\documentclass[12pt,preprint]{interact}
\usepackage[latin9]{inputenc}
\usepackage{color}
\usepackage{verbatim}
\usepackage{refstyle}
\usepackage{amsmath}
\usepackage{graphicx}
\PassOptionsToPackage{version=3}{mhchem}
\usepackage{mhchem}

\makeatletter

\AtBeginDocument{\providecommand\subsecref[1]{\ref{subsec:#1}}}
\AtBeginDocument{\providecommand\figref[1]{\ref{fig:#1}}}
\AtBeginDocument{\providecommand\tabref[1]{\ref{tab:#1}}}
\providecommand{\tabularnewline}{\\}
\RS@ifundefined{subsecref}
  {\newref{subsec}{name = \RSsectxt}}
  {}
\RS@ifundefined{thmref}
  {\def\RSthmtxt{theorem~}\newref{thm}{name = \RSthmtxt}}
  {}
\RS@ifundefined{lemref}
  {\def\RSlemtxt{lemma~}\newref{lem}{name = \RSlemtxt}}
  {}

\usepackage{siunitx}
\usepackage{mhchem}
\usepackage{pgfplots}
\usepackage[english]{babel}
\usepackage[sort&compress,numbers,merge,super]{natbib}
\usepackage{amsmath}
\usepackage{amssymb}
\usepackage{stmaryrd}
\usepackage{xfrac}
\renewcommand{\tabref}{\Tabref}
\renewcommand{\figref}{\Figref}

\pgfplotsset{compat=1.3}
\usetikzlibrary{patterns}
\usetikzlibrary{plotmarks}

\makeatother

\begin{document}
\title{Transition-Potential Coupled Cluster II: Optimization of the Core
Orbital Occupation Number}
\author{\name{Megan Simons$^\ast$\thanks{$^\ast$E-mail address: msimons@smu.edu}}
\name{Devin A. Matthews}\affil{Southern Methodist University, Dallas, TX 75275}}
\maketitle
\begin{abstract}
The issue of orbital relaxation in computational core-hole spectroscopy,
specifically x-ray absorption, has been a major problem for methods
such as equation-of-motion coupled cluster with singles and doubles
(EOM-CCSD). The transition-potential coupled cluster (TP-CC) method
is utilized to address this problem by including an explicit treatment
of orbital relaxation via the use of reference orbitals with a fractional
core occupation number. The value of the fractional occupation parameter
$\lambda$ was optimized for both TP-CCSD and XTP-CCSD methods in
an element-specific manner due to the differences in atomic charge
and energy scale. Additionally, TP-CCSD calculations using the optimized
parameters were performed for the K-edge absorption spectra of gas-phase
adenine and thymine. TP-CCSD reproduces the valence region well and
requires smaller overall energy shifts in comparison to EOM-CCSD,
while also improving on the relative position and intensities of several
absorption peaks.

\begin{keywords}Coupled cluster; NEXAFS; XPS; excited states\end{keywords}
\end{abstract}

\section{Introduction}

X-ray absorption spectroscopy (XAS) has been used for many years by
experimental chemists to study electronic and molecular structure,\cite{bokhovenXRayAbsorptionXRay2016,bergmannXRayFreeElectron2017,krausUltrafastXraySpectroscopic2018}
but theoretical calculations can often assist in interpreting spectra
as well as predicting spectra of unknown compounds. X-ray techniques
are important in addressing the challenges of understanding the structure
of complex molecules and understanding the behavior of these molecules.
Historically, theoretical chemists have predominately used density
functional theory (DFT) methods for XAS.\cite{normanSimulatingXraySpectroscopies2018}
However, techniques such as Transition-Potential DFT (TP-DFT) improve
on TD-DFT, and the systematic improvability of equation-of-motion
coupled cluster (EOM-CC) methods provides a significant motivation
for applying EOM-CC to XAS.

TP-DFT directly addresses the problem of orbital relaxation which
plagues even EOM-CC theory by construction a fractionally-occupied
reference state. On the other hand, within the EOM-CC framework the
inclusion of two-electron excitations in the excited state response
operator ($\hat{R}$ vector) allows EOM-CC to recover much of the
relaxation error, and triple excitations are sufficient to fully overcome
this problem.\cite{734c7d7e72234a548c61e39e075d7bd2} However, the
combination of fairly large residual errors of as much as 1--3 eV
at the doubles level and the excessive computational cost of triple
excitations calls for a novel solution.

We recently introduced a Transition-Potential Coupled Cluster (TP-CC)
method\cite{simonsTransitionpotentialCoupledCluster2021} in analogy
to TP-DFT, in hopes of reducing the orbital relaxation error. This
method utilizes the best parts of TP-DFT and EOM-CC and was able to
successfully reduce errors in both the absolute and relative (i.e.
empirically shifted) transition energies and intensities. This improvement
comes about due to a cancellation of errors between the destabilized
ground state and stabilized core-hole state which counterbalance the
overestimation of core excitation energies at the singles and doubles
level. The precise point of cancellation depends on the orbitals used,
however, with the total number of core electrons removed from the
core orbital, $\lambda$, as the controlling parameter. In this previous
work, which we will refer to as Paper I, we tested only $\lambda=1/2$
(in analogy to TP-DFT) and $\lambda=1/4$. Unlike TP-DFT, though,
there is no theoretical basis for such an a priori choice in TP-CC
and in fact other values may produce superior results. Additionally,
the optimal value of $\lambda$ may vary in an element-specific manner
due to the differences in atomic charge and energy scale.

In this work, we systematically investigate the optimal $\lambda$
parameter as a function of element for molecules containing first-row
elements CNOF, and finally apply TP-CC with the optimized parameters
to the example of the nitrogen, oxygen, and carbon K-edge spectra
of gas-phase nucleobases.

\section{Transition-Potential Coupled Cluster}

Equation-of-motion coupled cluster (or equivalently coupled cluster
linear response) theory\cite{monkhorstCalculationPropertiesCoupledcluster1977,mukherjeeResponsefunctionApproachDirect1979,sekinoLinearResponseCoupledcluster1984,kochCoupledClusterResponse1990,stantonEquationMotionCoupledcluster1993,comeauEquationofmotionCoupledclusterMethod1993}
computes excitation energies as eigenvalues of the transformed Hamiltonian,
$\bar{H}$,
\begin{align}
E_{CC} & =\langle0|\bar{H}|0\rangle\\
0 & =\langle P|\bar{H}|0\rangle\\
\omega_{m}\langle P|\hat{R}_{m}|0\rangle & =\langle P|[\bar{H},\hat{R}_{m}]|0\rangle
\end{align}
from which we can also easily incorporate the ground state with excitation
energy $\omega_{0}=0$ and response vector (right-hand eigenvector)
$\hat{R}_{0}=\hat{1}$. In EOM-CCSD the excitation space $\langle P|$
is restricted to singly- and doubly-excited determinants. Transition
intensities (oscillator strengths) are formally computed as residues
of the excitation poles, but we adopt the simpler and typically very
accurate expectation value formalism,
\begin{align}
f_{m}(\text{EOM-CC}) & =\frac{2m_{e}\omega_{m}}{3\hbar^{2}}\sum_{\alpha=x,y,z}M_{m,\alpha}\\
M_{m,\alpha} & =\langle0|\hat{L}_{0}\bar{\mu}_{\alpha}\hat{R}_{m}|0\rangle\langle0|\hat{L}_{m}\bar{\mu}_{\alpha}\hat{R}_{0}|0\rangle
\end{align}
where $\bar{\mu}_{\alpha}$ is the similarity-transformed electronic
dipole moment operator along the $\alpha$ Cartesian axis and $\hat{L}_{m}$
are the left-hand eigenvectors of $\bar{H}$.

In order to address the issue of orbital relaxation, we adopt a non-standard
reference state. We are motivated by TP-DFT theory,\cite{huDensityFunctionalComputations1996,trigueroCalculationsNearedgeXrayabsorption1998,trigueroSeparateStateVs1999,michelitschEfficientSimulationNearedge2019}
which is an approximation to Slater's Transition State (TS) method,
which in turn is ultimately derived from $\Delta$Kohn-Sham (or $\Delta$DFT).
Essentially, TP-DFT allows one to approximate a direct energy difference
between two states differing by the excitation of an electron from
one orbital to another, by an orbital energy difference obtained with
a fractional occupation,
\begin{align}
\omega_{\Delta KS} & =E_{f}-E_{i}\nonumber \\
 & =\int_{0}^{1}\left[\epsilon_{2}(\lambda)-\epsilon_{1}(\lambda)\right]d\lambda\nonumber \\
 & \approx\epsilon_{2}(1/2)-\epsilon_{1}(1/2)\nonumber \\
 & =\omega_{TP-DFT}\label{eq:tp-dft}
\end{align}
The choice of a half-electron occupation is justified by the trapezoidal
rule of numerical integration. In TP-CCSD, we also compute a fractionally-occupied
reference determinant (using a constrained variational optimization),
but then reoccupy the orbitals according to the Aufbau principle for
use in a non-Hartree--Fock EOM-CCSD calculation. While standard EOM-CCSD
typically overestimates core-hole excitation and ionization energies
by 1--3 eV,\cite{734c7d7e72234a548c61e39e075d7bd2,simonsTransitionpotentialCoupledCluster2021}
TP-CCSD largely corrects this error by combining a partial destabilization
of the ground state due to the use of non-HF orbitals and a partial
stabilization of the core-hole state due to the explicit inclusion
of core-hole relaxation. The balance between these effects is controlled
by the $\lambda$ parameter as defined in (\ref{eq:tp-dft}), defining
a family of methods TP-CCSD($\lambda$). Note however that in order
to maintain a spin-restricted reference state we \textcolor{black}{de-occupy}
both the $\alpha$ and $\beta$ core orbitals by $\lambda/2$ electrons
each. Importantly, this guarantees that the excited state is a spin
eigenfunction. We also defined an XTP-CC class of methods which differ
by promoting electrons from the core to the LUMO instead of ionizing
them. This maintains a charge-neutrality while optimizing the orbitals
and also includes some stabilization of the LUMO (often a $\pi^{*}$
orbital).

In Paper I we showed that TP-CCSD($\sfrac{1}{2}$) was highly effective
in reducing errors in both transition energies and intensities relative
to EOM-CCSD. However, $\lambda=1/2$ is by no means guaranteed to
provide the lowest errors. In fact, we expect the optimal value of
lambda to vary depending on the atomic number, as the size of the
orbital relaxation error grows with energy scale and hence atomic
charge. As TP-CCSD is essentially a ``normal'' EOM-CCSD calculation
with a different choice of orbitals, both the calculation of core-hole
states via the core-valence separation (CVS)\cite{corianiCommunicationXrayAbsorption2015}
and the calculation of transition intensities are identical to EOM-CCSD.
We use the CVS for all EOM-CC and TP-CC calculations in this work.

\section{Computational Details}

The (X)TP-CCSD($\lambda$) methods were implemented via a combination
of the Psi4\cite{smithPSI4OpensourceSoftware2020} and CFOUR\cite{matthewsCoupledclusterTechniquesComputational2020}
program packages. We used a locally-modified version of the PSIXAS
plugin\cite{ehlertPSIXASPsi4Plugin2020a} for Psi4 to generate fractional
core-hole or core-excited orbitals at the B3LYP level of theory. 

The test set used and methodology are the same as in Paper I, except
where noted below. We performed calculations with values of $\lambda$
from 0.25 to 0.75 in steps of 0.025 (21 points total). We use the
same full CVS-EOM-CCSDT benchmark as in Paper I, except in the calculations
of nucleobase spectra where we compare to experimental data. In the
latter case, we include an estimate of relativistic effects as +0.38
eV for the oxygen K-edge, +0.21 eV for the nitrogen K-edge, and +0.10
eV for the carbon K-edge. The relativistic contributions are very
weakly-dependent on the chemical environment for first-row K-edges.\cite{734c7d7e72234a548c61e39e075d7bd2}
Note that in Paper I we selected four excited states for each edge
on the basis of dominant single-excitation character and non-negligible
oscillator strengths. This is critical as doubly-excited states are
poorly treated at the singles and doubles level. We changed a small
number of state selections in this work to avoid states which mix
strongly with doubly-excited states for some values of $\lambda$,
although certain unavoidable crossings remain (see \subsecref{Oscillator-Strengths}).

Variation of errors across the test set are quantified by computing
the mean absolute error (MAE) across the entire test set, except for
a small number of calculations which did not converge for large values
of $\lambda$. The full data set is available in the electronic Supporting
Information file.

\section{Results and Discussion}

\subsection{Excitation and Ionization Energies}

\begin{figure}
\includegraphics[scale=0.5]{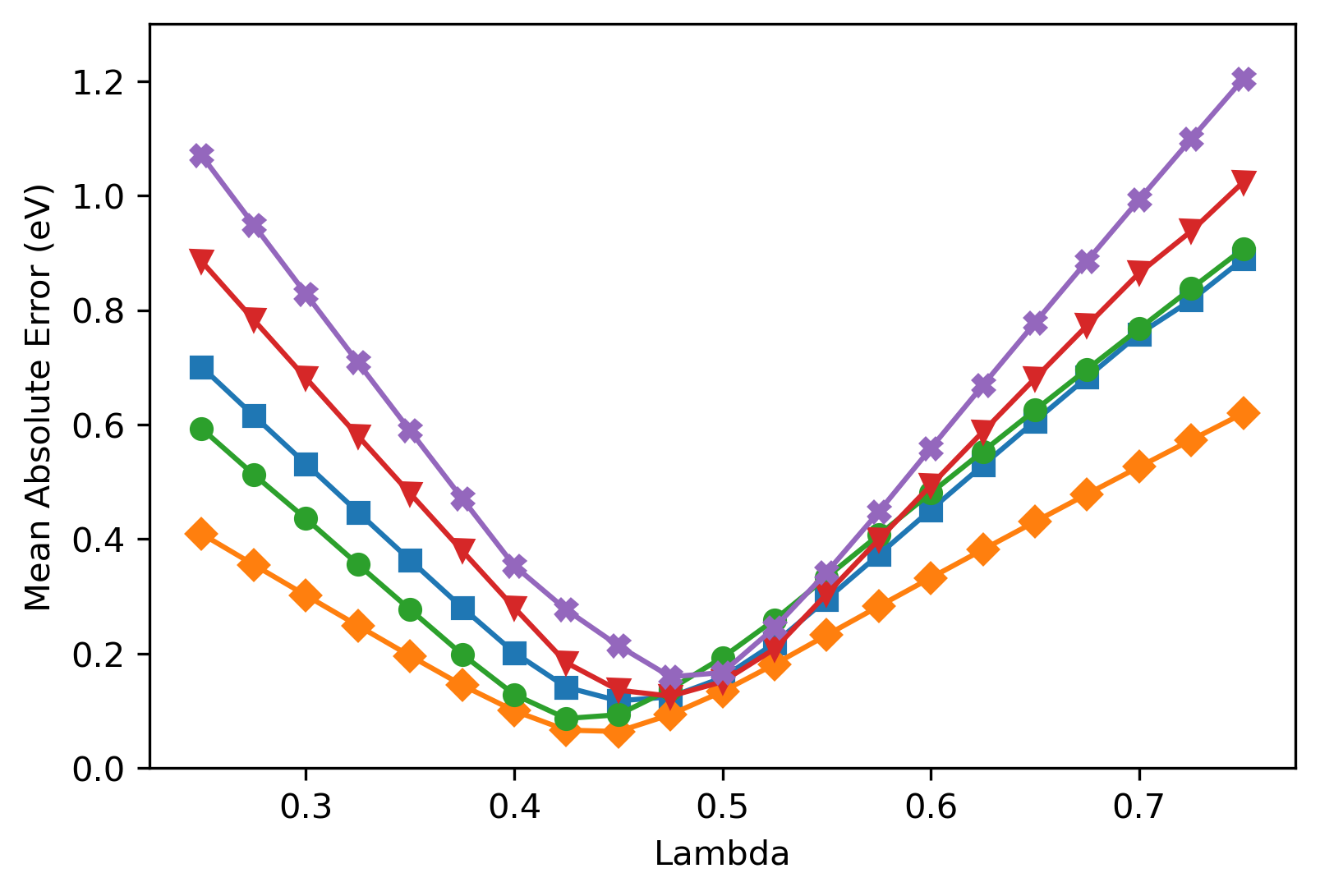}\includegraphics[scale=0.5]{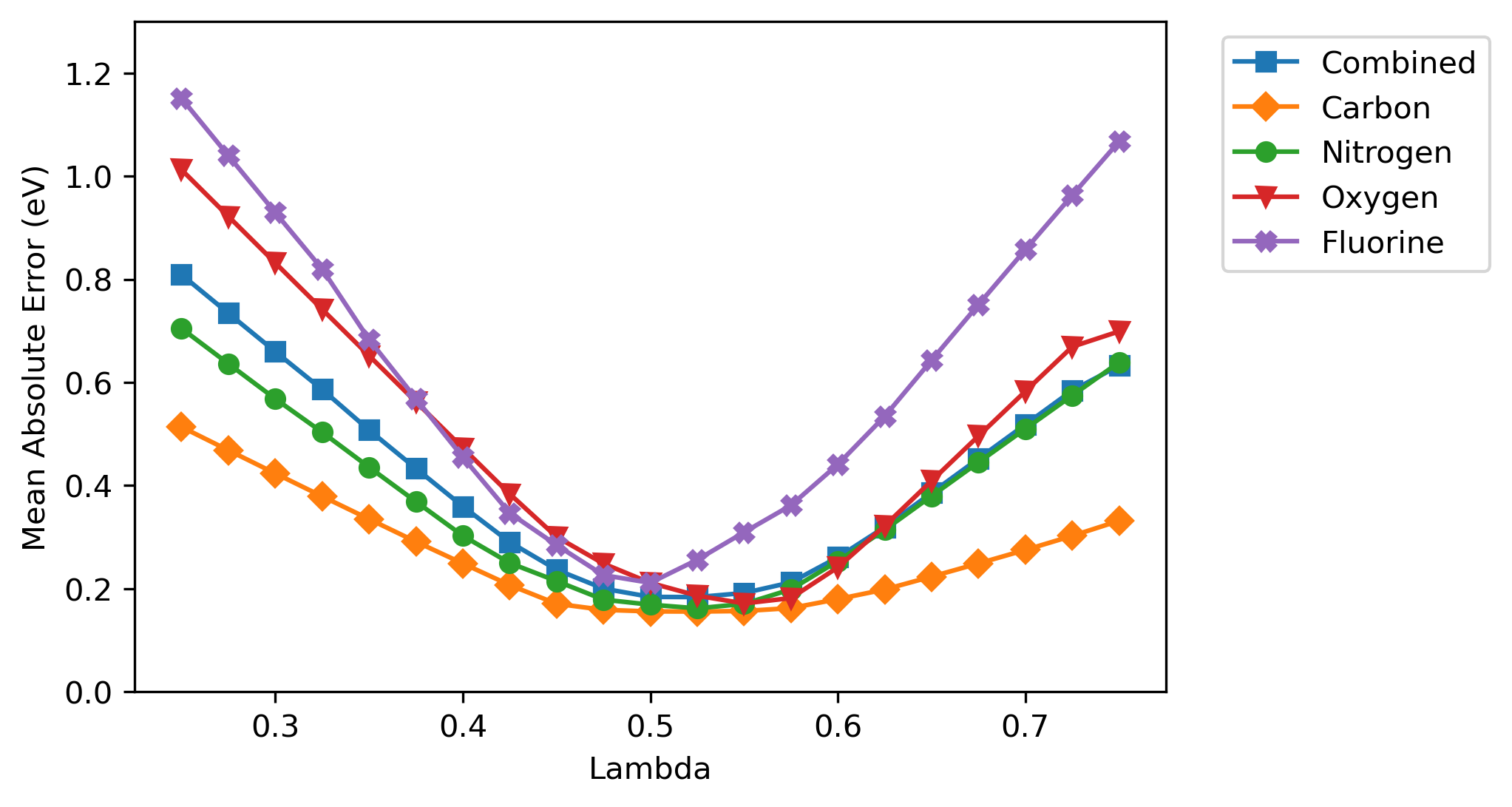}

\caption{\label{fig:absolute}TP-CCSD (left) and XTP-CCSD (right) mean absolute
error distributions for absolute (vertical) core excitation energies.
Errors specific to the K-edges of elements C--F are reported separately,
and a combined MAE value is also included.}
\end{figure}
\begin{figure}
\includegraphics[scale=0.5]{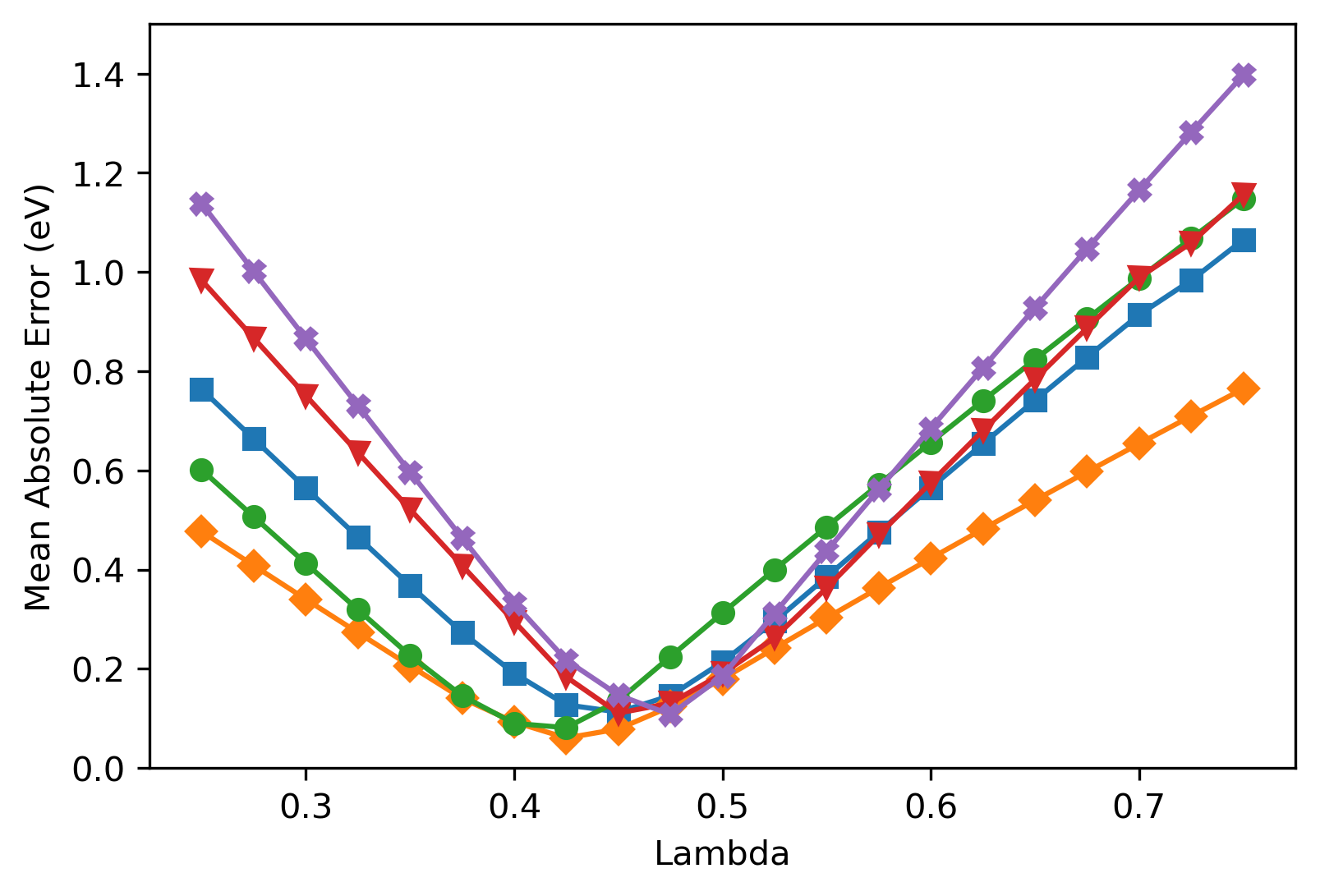}\includegraphics[scale=0.5]{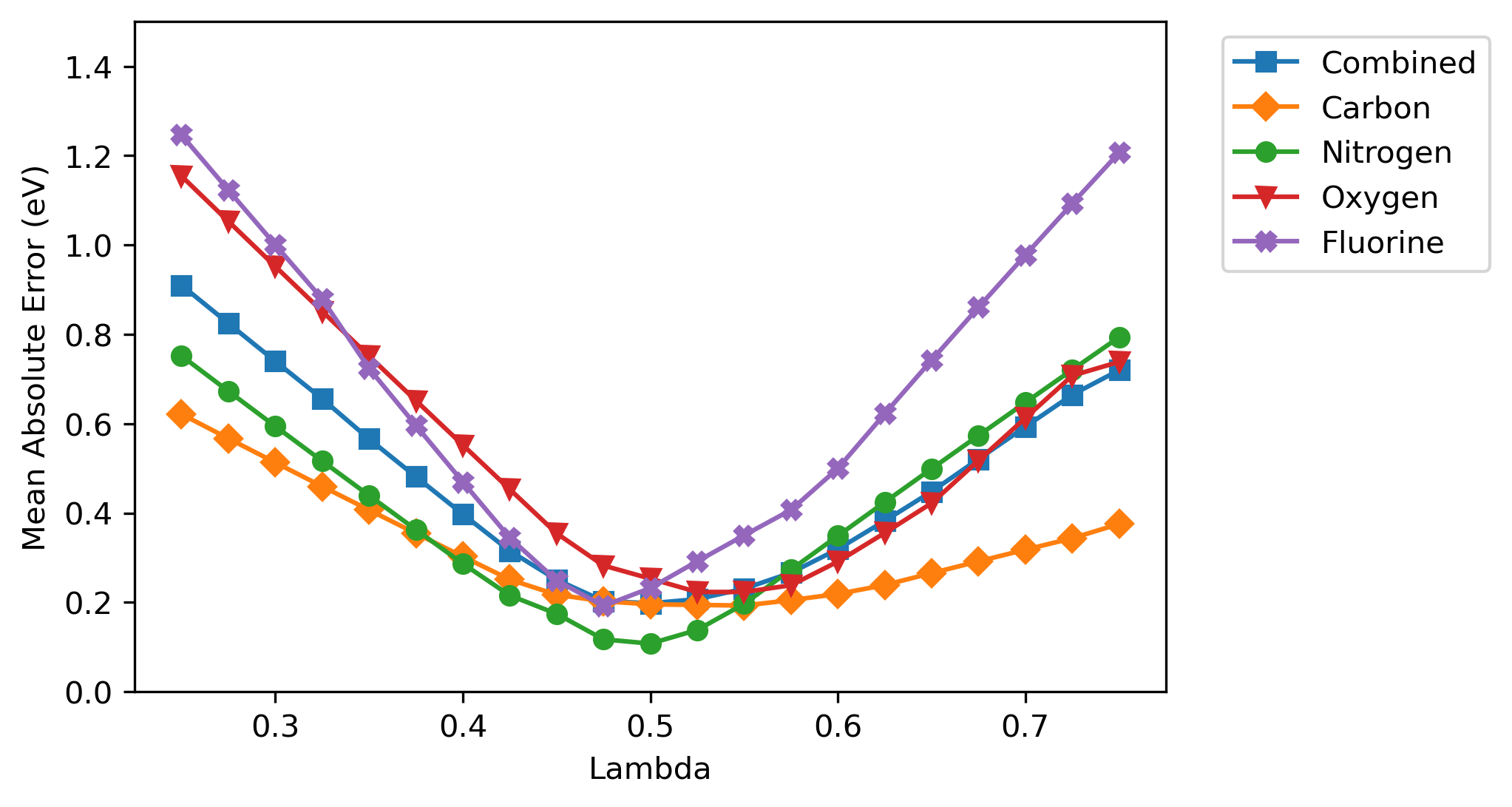}

\caption{\label{fig:ip}TP-CCSD (left) and XTP-CCSD (right) mean absolute error
distributions for absolute (vertical) core ionization energies. Errors
specific to the K-edges of elements C--F are reported separately,
and a combined MAE value is also included.}
\end{figure}
\begin{figure}
\includegraphics[scale=0.5]{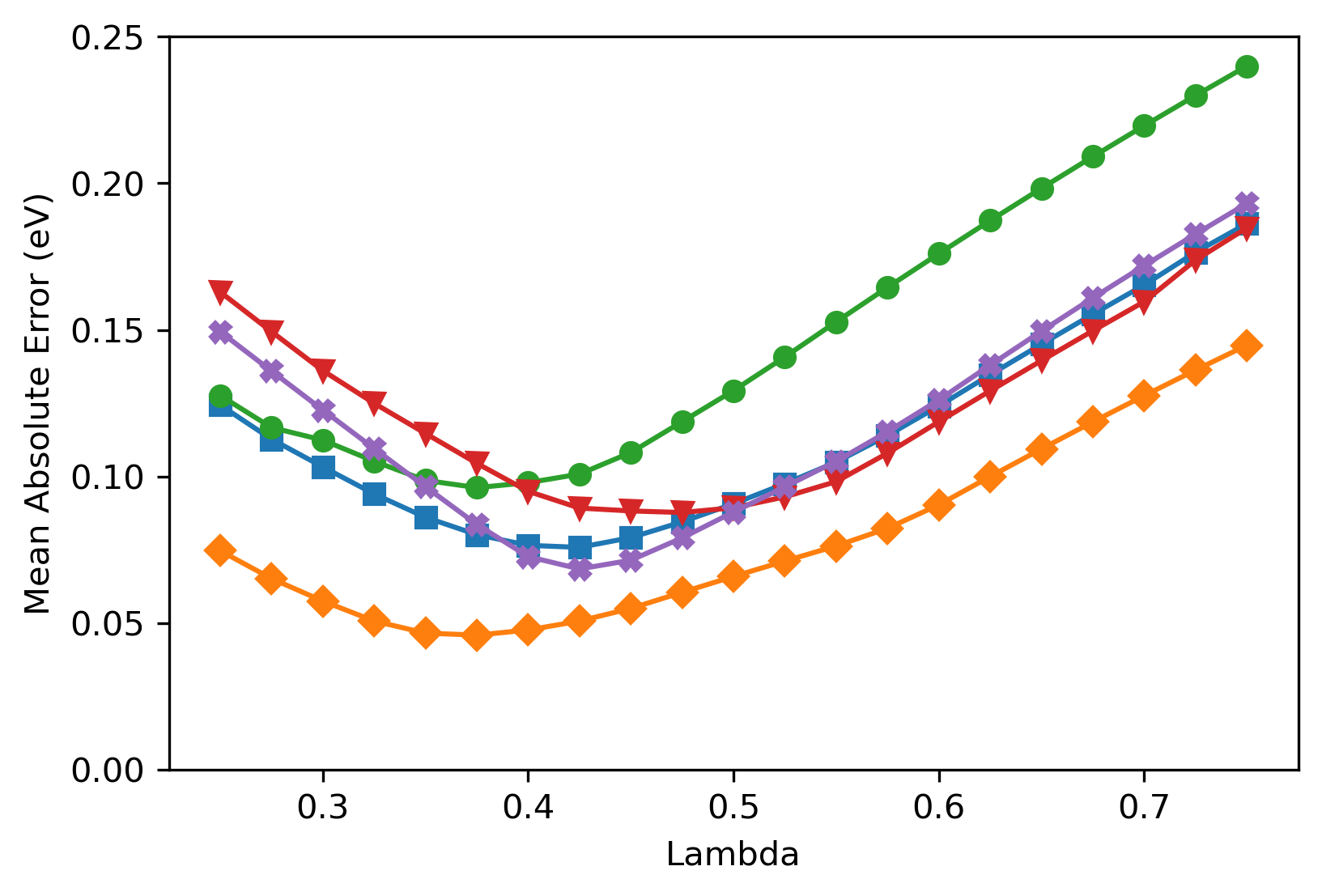}\includegraphics[scale=0.5]{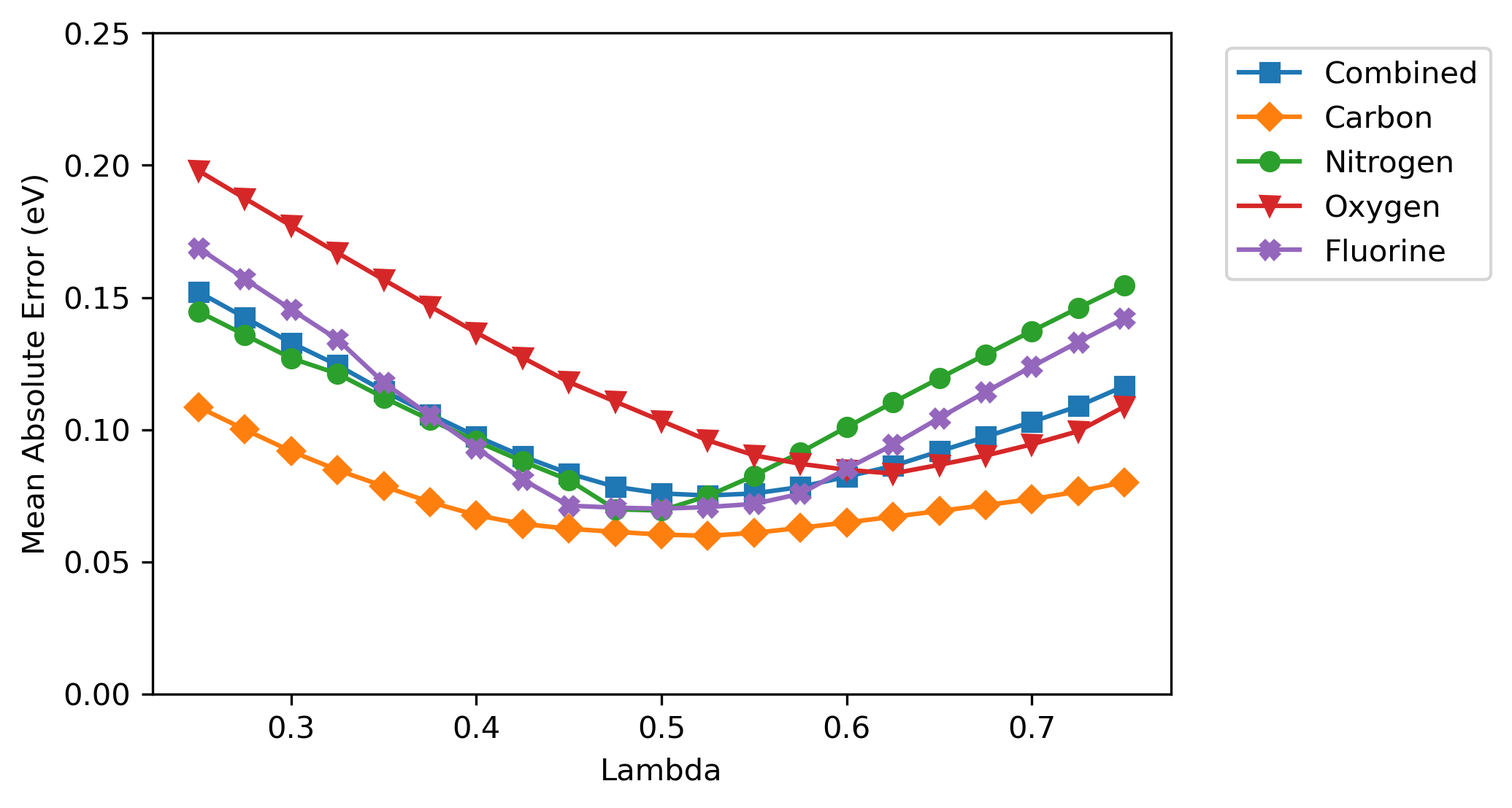}

\caption{\label{fig:rel}TP-CCSD (left) and XTP-CCSD (right) mean absolute
error distributions for relative core excitation energies (see text).
Errors specific to the K-edges of elements C--F are reported separately,
and a combined MAE value is also included.}
\end{figure}

The distribution of ``absolute'' (i.e. unmodified vertical) excitation
energy deviations from CVS-EOM-CCSDT is depicted in \figref{absolute}.
The MAE distribution of computed ionization potentials is depicted
in \figref{ip}. Finally, the MAE distribution of ``relative'' excitation
energies is depicted in \figref{rel}. The relative deviations are
determined from excitation energies measured from the corresponding
ionization edge (this is essentially a shift of the entire spectrum---note
that the shift is method- and $\lambda$-dependent, and is applied
before computing the MAE). Since each method should make similar errors
in the ionization potential energies and excitation energies, the
relative errors should be smaller due to error cancellation. A similar
shift is commonly applied when comparing to experimental data. 

In Paper I, the choice of $\ensuremath{\lambda=\sfrac{1}{2}}$ was
shown to be a reasonable first-order estimate of the optimal error-cancellation
point for TP-CC methods. From Figs. \ref{fig:absolute} and \ref{fig:ip}
we can see that there is indeed a fairly sharp minimum in the mean
absolute energy as $\lambda$ is varied, with somewhat sharper minima
for TP-CCSD than for XTP-CCSD. However, in many cases the optimal
$\lambda$ value is significantly different from 0.5. For example,
the optimal value for nitrogen 1s excitations with TP-CCSD is approximately
0.425, while the best value for oxygen 1s excitations with XTP-CCSD
is approximately 0.575. The optimal value of $\lambda$ also depends
strongly on the atomic number of the 1s core orbital. Looking at the
``combined'' MAEs shows that the blindly-averaged results do not,
in many cases, obtain errors as low as when considering elements independently
nor predict the best value of $\lambda$ for any particular edge.
Due to the sharpness of the error distributions, the simple choice
of $\lambda=1/2$ may result in errors twice as large or more compared
to a ``tuned'' value. While we do not advocate optimizing the value
of $\lambda$ for any specific spectrum, it does seem clear that large
accuracy gains can be obtained by making a more informed and element-specific
choice of the $\lambda$ parameter.

The relative excitation energies, owing to a significant cancellation
of errors between the excitation and ionization energy calculations,
do not show a clear trend with atomic number, although individual
MAEs are significantly lower than for absolute energies. As with excitation
energies, evaluating relative ionization energies, specifically ionization
``chemical shifts''\cite{liuBenchmarkCalculationsKEdge2019} relative
to a standard species, may similarly reduce the ionization potential
errors. Despite the lack of overall trends, there is still a strong
effect of atomic number on the optimal value of $\lambda$. In fact,
the deviation of the optimal value from our previous choice of 0.5
is even more pronounced, with carbon and nitrogen K-edges benefiting
most from a value of $\lambda$ as low as 0.35. The combined TP-CCSD
relative excitation energy results do show that a ``global'' choice
of 0.425 results in fairly good treatment regardless of atomic number,
with error increasing over the optimal $\lambda$ values by only 20\%
or so. However, choosing slightly better element-specific values of
$\lambda$ is trivial to do and results in the best predictions of
core excitation spectra. We will defer choosing a recommended set
of $\lambda$ values until \subsecref{Oscillator-Strengths}.

Splitting up the test set based on the atomic edge (C 1s through F
1s) reveals several interesting patterns in the absolute excitation
and ionization energy results. Firstly, the errors at low values of
$\lambda$ (and similarly in standard CVS-EOM-CCSD) are strongly dependent
on atomic number and increase proportionally. This reflects the increasing
energy scale of the 1s orbital energies. Additionally, the optimal
value of $\lambda$ for TP-CCSD increases with increasing atomic number.
This indicates that a larger amount of core-hole character must be
included in the reference orbitals in order to cancel out the orbital
relaxation error. As the orbital relaxation contribution increases
along with the absolute energy scale, this trend is not unexpected.
However, this trend is completely reversed in the XTP-CCSD results,
with the optimal value of $\lambda$ now decreasing with atomic number.
As XTP-CC places a fractional electron into the LUMO, this must indicate
a stronger stabilization of the LUMO with atomic number. This may
derive from the increasing electronegativity of the elements along
the first row of the periodic table, which alters the shape (in particular
the polarization) of the LUMO, which is typically a $\pi^{*}$ orbital.
These clear trends disappear in the case of relative excitation energies
due to cancellation of the overall trends in the absolute energies.

\begin{table}
\setlength\tabcolsep{2pt}%
\begin{tabular}{|c|c|c|c|c|c|}
\hline 
 & Carbon 1s & Nitrogen 1s & Oxygen 1s & Fluorine 1s & Combined\tabularnewline
\hline 
Abs. EE (eV) & 1.07 (16.8) & 1.46 (16.9) & 1.90 (15.1) & 2.19 (13.8) & 1.57 (13.4)\tabularnewline
\hline 
Abs. IP (eV) & 1.29 (21.3) & 1.64 (20.1) & 2.10 (19.0) & 2.38 (22.0) & 1.76 (15.6)\tabularnewline
\hline 
Rel. EE (eV) & 0.390 (8.5) & 0.471 (4.9) & 0.524 (6.5) & 0.574 (8.4) & 0.475 (6.3)\tabularnewline
\hline 
Abs. $f$ ($\times100$) & 2.43 (18.0) & 2.12 (14.3) & 1.95 (20.8) & 1.33 (17.1) & 2.04 (15.6)\tabularnewline
\hline 
Rel. $f$ (\%) & 24.6\% (16.4) & 15.6\% (10.8) & 20.5\% (34.4) & 29.5\% (15.4) & 22.4\% (16.0)\tabularnewline
\hline 
\end{tabular}

\caption{\label{tab:Mean-absolute-error}Mean absolute error values for standard
CVS-EOM-CCSD core excitation energies (EE), core ionization energies
(IP), and oscillator strengths ($f$). Errors are broken down categorically
as in Figs. 1--5; in each case the values in parenthesis indicate
the ratio of the CVS-EOM-CCSD error to the best value obtainable by
TP-CCSD. Note that the combined TP-CCSD MAEs used for comparison are
chosen based on the minimum of the ``combined'' curve in each figure,
and not from a combination of the individual MAEs used for the element-specific
comparisons, although this approach would lead to lower errors.}
\end{table}

The deviations corresponding to Figs. 1--5 for standard CVS-EOM-CCSD
are reproduced in \tabref{Mean-absolute-error}, which of course do
not depend on a $\lambda$ parameter. For ease of comparison, we have
included the ratio of the CVS-EOM-CCSD MAEs to the best TP-CCSD values
(i.e. the minima of the curves in the figures) in parentheses. From
these results, we can see that TP-CCSD is easily capable of reducing
the size of the orbital relaxation error in vertical core-excitation
and core-ionization energies by approximately a factor of 15 (and
as high as a factor of 22) compared to a standard EOM calculation.
Given that TP-CCSD incurs nearly zero additional computational cost,
it seems that TP-CCSD is highly preferable for routine calculations.
Errors in relative excitation energies (intra-edge) are also significantly
reduced by a factor of 5 to 8. This leads to errors in the line positions
of approximately 100 meV, which is on par with or smaller than the
typical instrument broadening. Equally as important as the line positions,
the errors oscillator strengths (either relative or absolute, discussed
in more detail in \subsecref{Oscillator-Strengths}) are also reduced
by at least a factor of 15 compared to CVS-EOM-CCSD. Relative oscillator
strengths are correct in TP-CCSD to within 2--3\% (compared to our
CVS-EOM-CCSDT reference), which enables a significantly more reliable
basis for experimental assignment.

\subsection{Oscillator Strengths\label{subsec:Oscillator-Strengths}}

Because TP-CC is computationally identical to a standard EOM-CC calculation
apart from the choice of orbitals, it is simple to compute oscillator
strengths using the expectation value formalism. By extension, it
is also trivial to compute other important excited state properties
such as excited state dipole moment, natural transition orbitals,
and extent of diffusion ($\langle r^{2}\rangle$ values).

\begin{figure}
\includegraphics[scale=0.5]{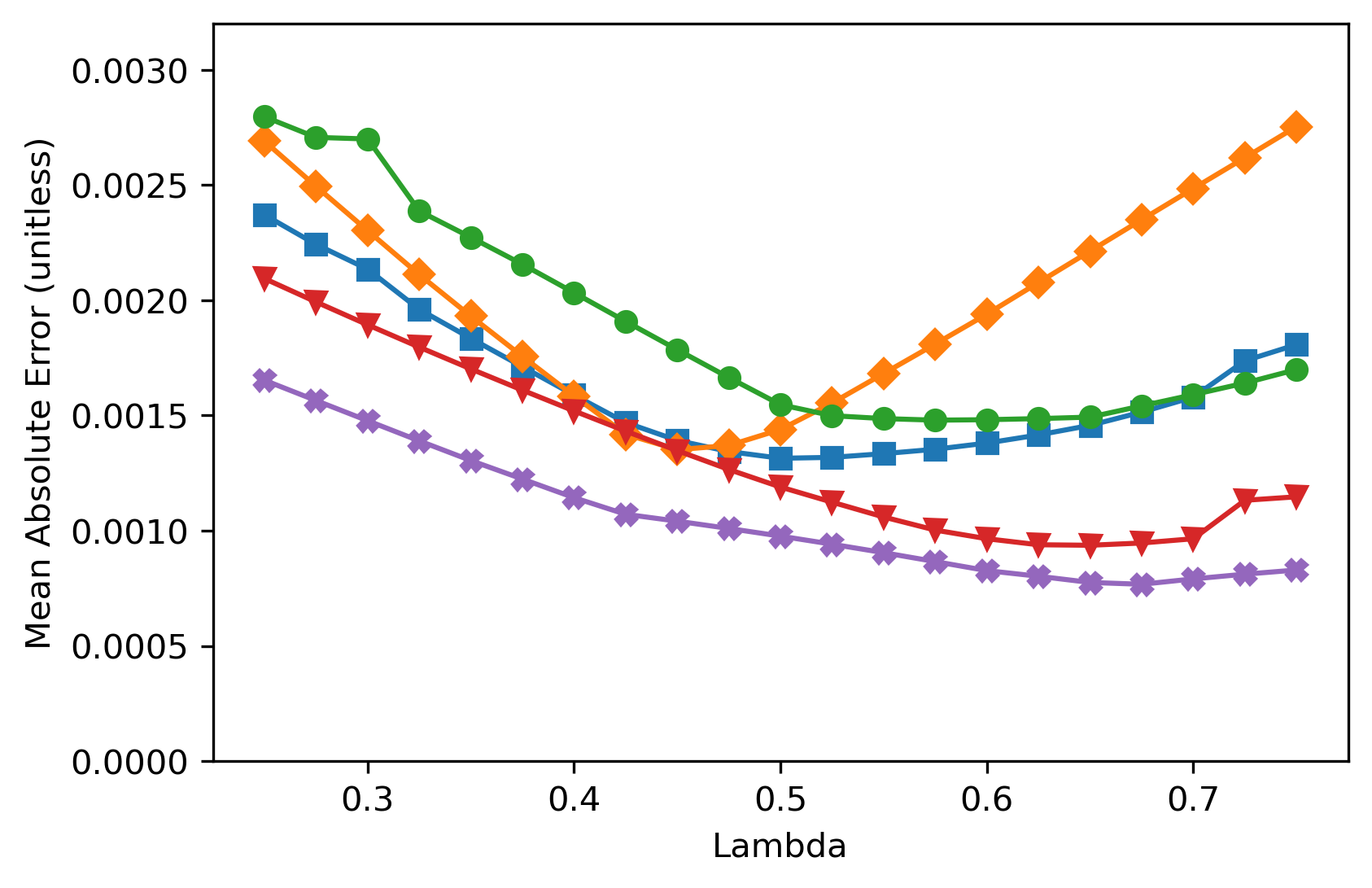}\includegraphics[scale=0.5]{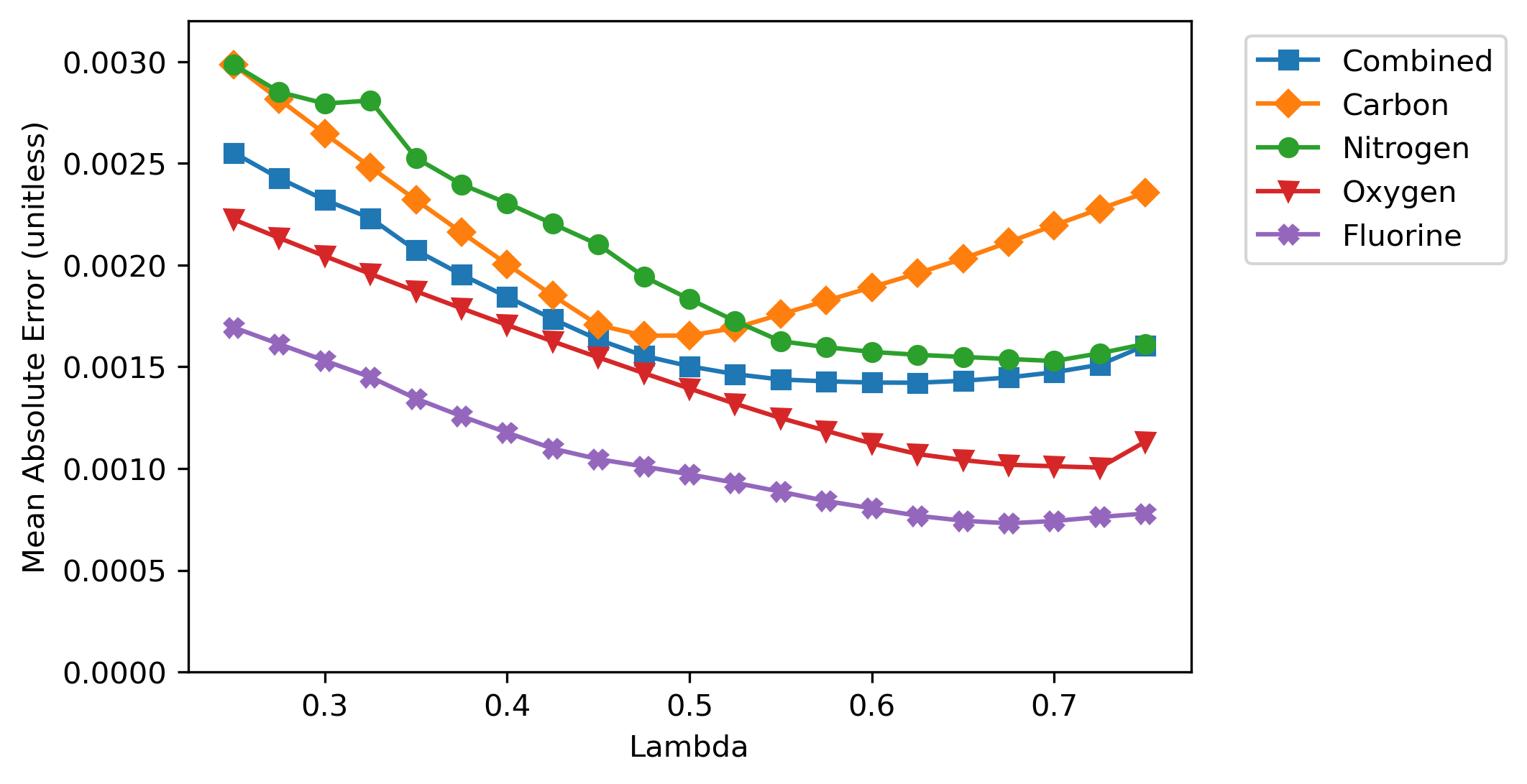}

\caption{\label{fig:absolute-int}TP-CCSD (left) and XTP-CCSD (right) mean
absolute error distributions for absolute (dimensionless) oscillator
strengths. Errors specific to the K-edges of elements C--F are reported
separately, and a combined MAE value is also included.}
\end{figure}
\begin{figure}
\includegraphics[scale=0.5]{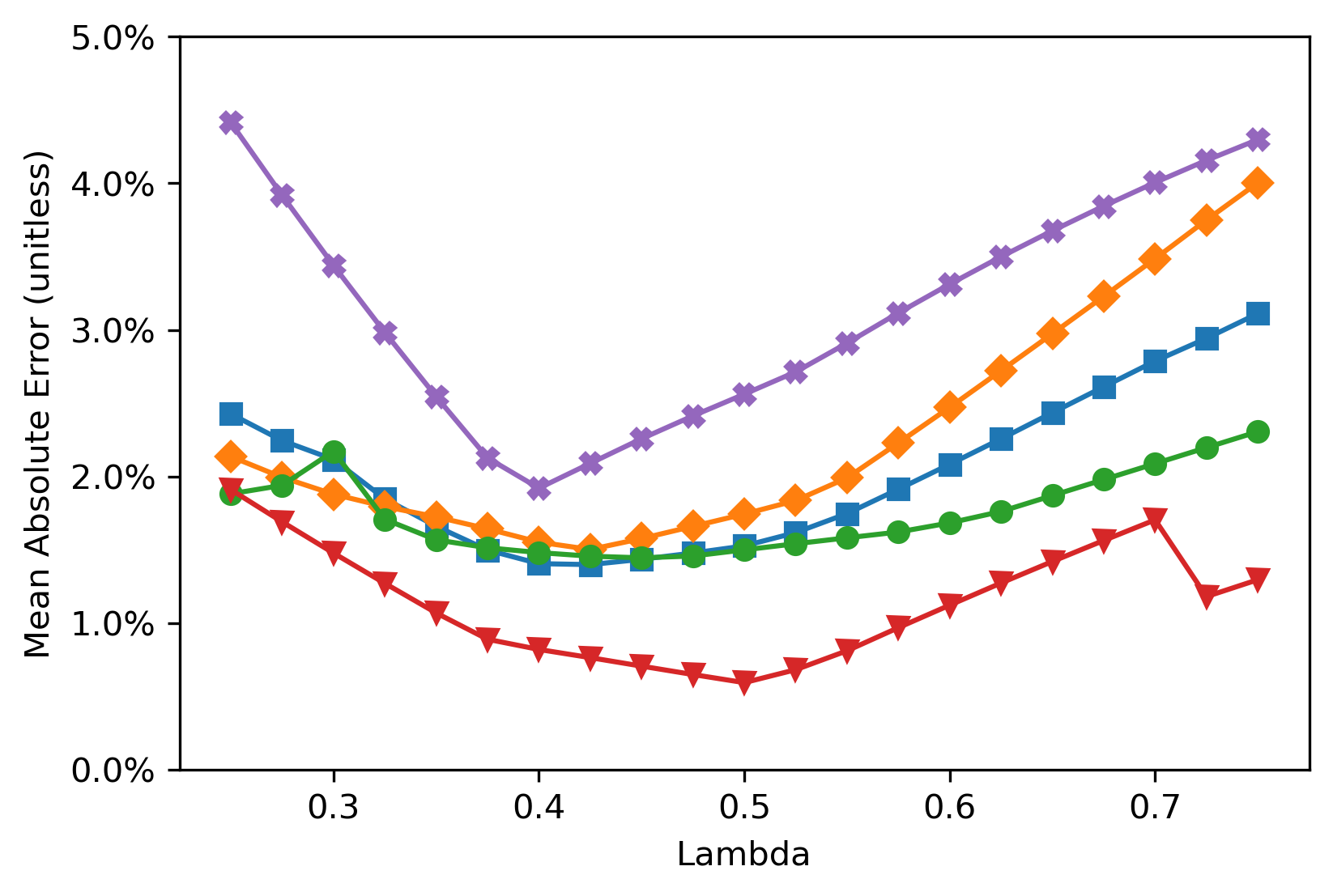}\includegraphics[scale=0.5]{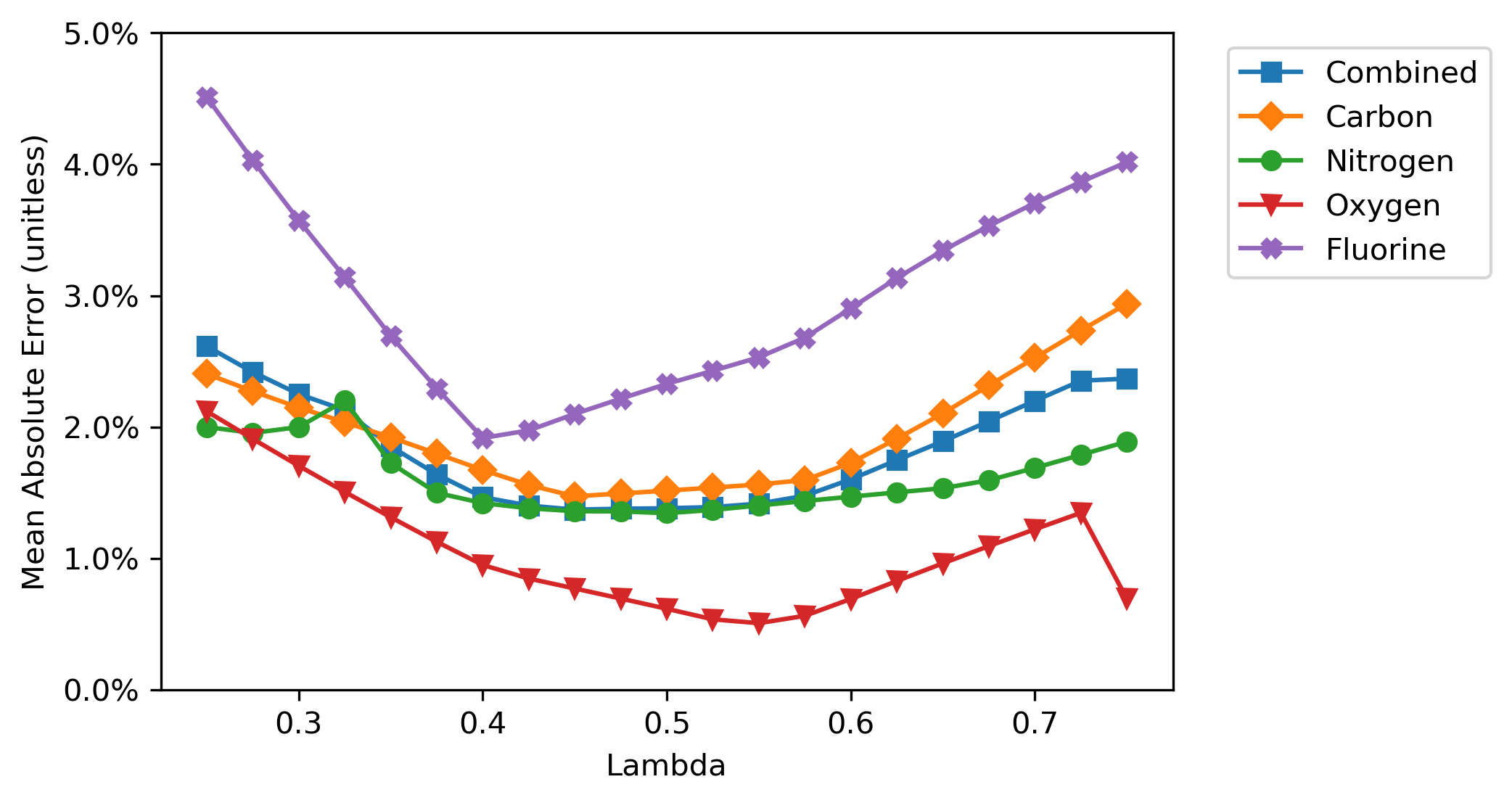}

\caption{\label{fig:rel-int}TP-CCSD (left) and XTP-CCSD (right) mean absolute
error distributions for relative oscillator strengths (see text).
Errors specific to the K-edges of elements C--F are reported separately,
and a combined MAE value is also included.}
\end{figure}

We calculated dimensionless oscillator strengths for all transitions
and compared to benchmark CVS-EOM-CCSDT values from Paper I. The mean
absolute deviations in the oscillator strengths, defined in analogy
to the absolute excitation and ionization energy deviations, are depicted
in \figref{absolute-int} as a function of $\lambda$. In addition,
we have computed the mean absolute deviations of relative oscillator
strengths. In this case we separately normalized the spectrum of each
edge such that the most intense transition has unit strength; the
statistics for these relative deviations are depicted in \figref{rel-int}
as percentages.

Due to a lack of convergence of either the ground-state CCSD or EOM-CCSD
equations, several data points are not available for $\lambda=0.725$
and $\lambda=0.75$. These data points were omitted from the MAE statistics,
but the effect is clearly discernible in Figs. \ref{fig:absolute-int}
and \ref{fig:rel-int} due to the large variations in oscillator strengths
across the test set. Additionally, an irregularity in the nitrogen
1s MAE curves is clearly visible, with a sharp local maximum near
$\lambda=0.3$. This is due to mixing with a dark doubly-excited state
near this value of $\lambda$, which is not present in the reference
CVS-EOM-CCSDT calculations.

Both MAE distributions for absolute and relative oscillator strengths
have broad and rather unstructured minima, with the exception of the
carbon K-edges. Additionally, the smaller overall MAE for fluorine
can probably be attributed to the low number of data points for fluorine
containing molecules. While an extension of the test set to better
represent fluorine K-edge could improve the statistical relevance
of the MAE curves, it seems clear that, in general, the oscillator
strengths are less dependent on a specific value of $\lambda$ compared
to the excitation and ionization energies. The increase in error for
the ``worst'' value of $\lambda$ tested compared to the optimal
value is less than a factor of 2 in all cases. Additionally, any ``reasonable''
value of $\lambda$ leads to a significantly reduced error, as demonstrated
in \tabref{Mean-absolute-error}. While we did not systematically
examine $0<\lambda<0.25$, the disconnect between the shallowness
of the MAE curves in the $0.25\le\lambda\le0.75$ region and the large
reduction of error relative to CVS-EOM-CCSD (equivalent to $\lambda=0$)
necessitates a ``bathtub''-like shape of the MAE curve, with a sharp
decrease at low values of $\lambda$ followed by a relatively flat
region.

Because the accuracy of the oscillator strengths depends only weakly
on the choice of $\lambda$, and because both the absolute and relative
errors are very low in any case, we base our recommended values of
$\lambda$ on the minimum MAE points in \figref{rel}. The absolute
energy errors are typically less important than the relative shifts,
for example when assigning experimental spectra. The recommended values
for both TP- and XTP-CCSD calculations are presented in \tabref{recommended}.

\subsection{K-edge Absorption Spectra: Canonical Nucleobases\label{subsec:Nucleobases}}

\begin{table}
\begin{tabular}{|c|c|c|c|c|c|}
\hline 
 & Carbon 1s & Nitrogen 1s & Oxygen 1s & Fluorine 1s & Combined\tabularnewline
\hline 
TP-CCSD & 0.35 & 0.375 & 0.475 & 0.425 & 0.425\tabularnewline
\hline 
XTP-CCSD & 0.475 & 0.5 & 0.625 & 0.5 & 0.525\tabularnewline
\hline 
\end{tabular}

\caption{\label{tab:recommended}Recommended values of $\lambda$ for TP-CCSD
and XTP-CCSD calculations of various K-edges. The ``combined'' values
are most suitable for a general, non-element-specific choice.}
\end{table}

As an additional benchmark, we computed carbon, nitrogen, and oxygen
K-edge spectra of adenine and thymine using TP-CCSD and the recommended
values of $\lambda$ (\tabref{recommended}), and compared the simulated
absorption spectra to available experimental gas-phase data\cite{plekanTheoreticalExperimentalStudy2008}
and standard CVS-EOM-CCSD calculations. The molecular geometries were
taken from Ref.~\citenum{leeEnergeticsVibrationalSignatures2018},
which are optimized at the M06-2X/aug-cc-pVTZ level. Vertical x-ray
absorption spectra were computed the same way as our previous TP-CCSD
calculations, utilizing a combination Psi4 and CFOUR, but we used
the more economical 6-311++G{*}{*} basis set. This basis set has shown
to be quite accurate for its size,\cite{734c7d7e72234a548c61e39e075d7bd2}
although we did not decontract the core orbital(s) as suggested in
Ref.~\citenum{sarangiBasisSetSelection2020}.

The final spectra are obtained by summing spectra from separate calculations
restricted to each C 1s, O 1s, or N 1s orbital in turn. Because of
the orbital-specific nature of the CVS, a similar procedure must be
followed in typical CVS-EOM-CC implementations, although the development
version of CFOUR supports mixing multiple edges in a standard CVS-EOM-CC
calculation. The calculation of each edge included the lowest 10 excited
states. We apply Lorentzian broadening with a half-width half-maximum
of 0.2 eV in all cases. %
The excited state energies are shifted so that the first peak in the
computational spectra lines up with the first peak in the experimental
spectra, as is customary. The shifts reported for each computational
spectra are the raw shift of the curve minus the estimated relativistic
effects for each K-edge.

\begin{figure}
\includegraphics[scale=0.65]{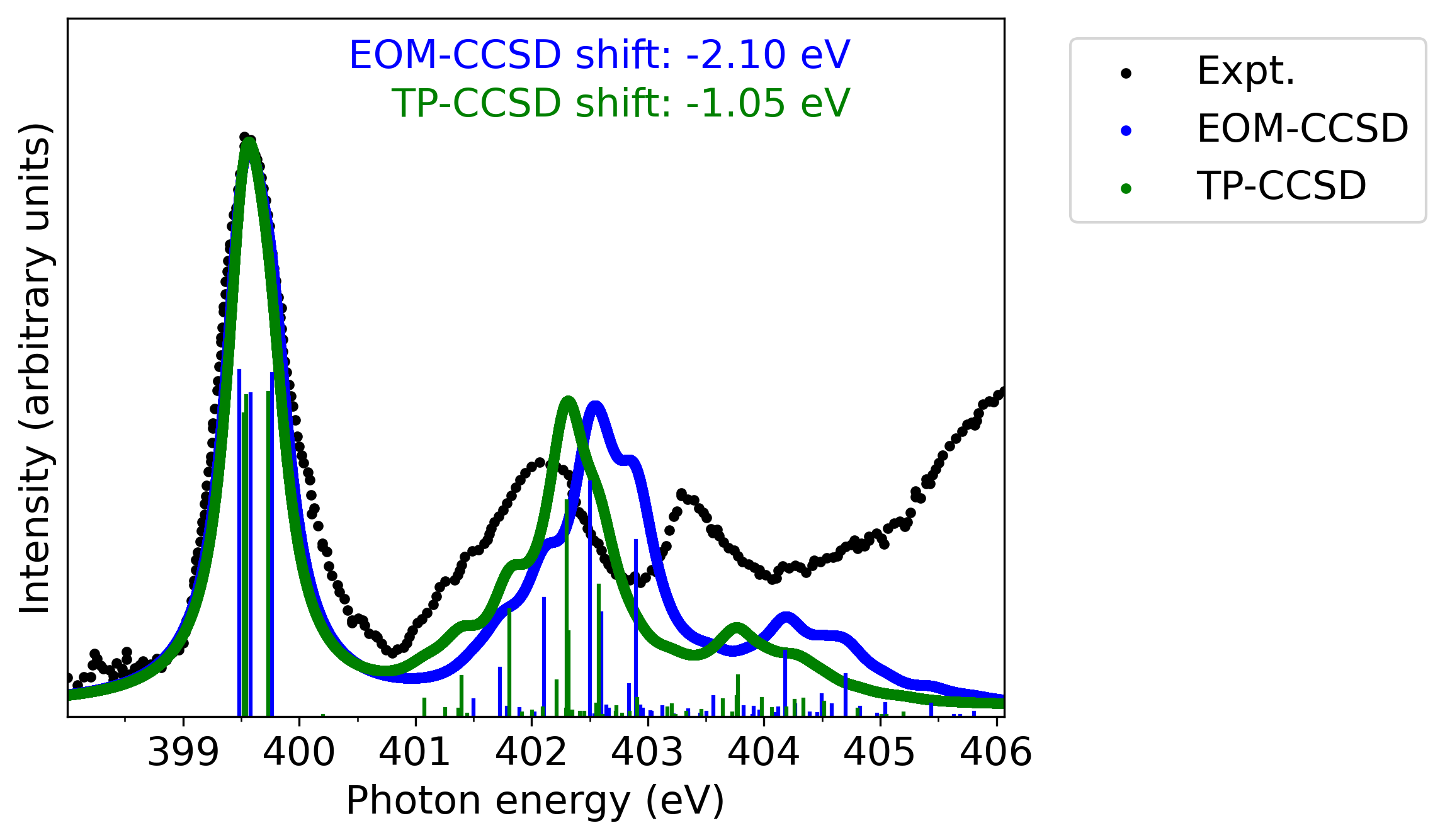}

\includegraphics[scale=0.65]{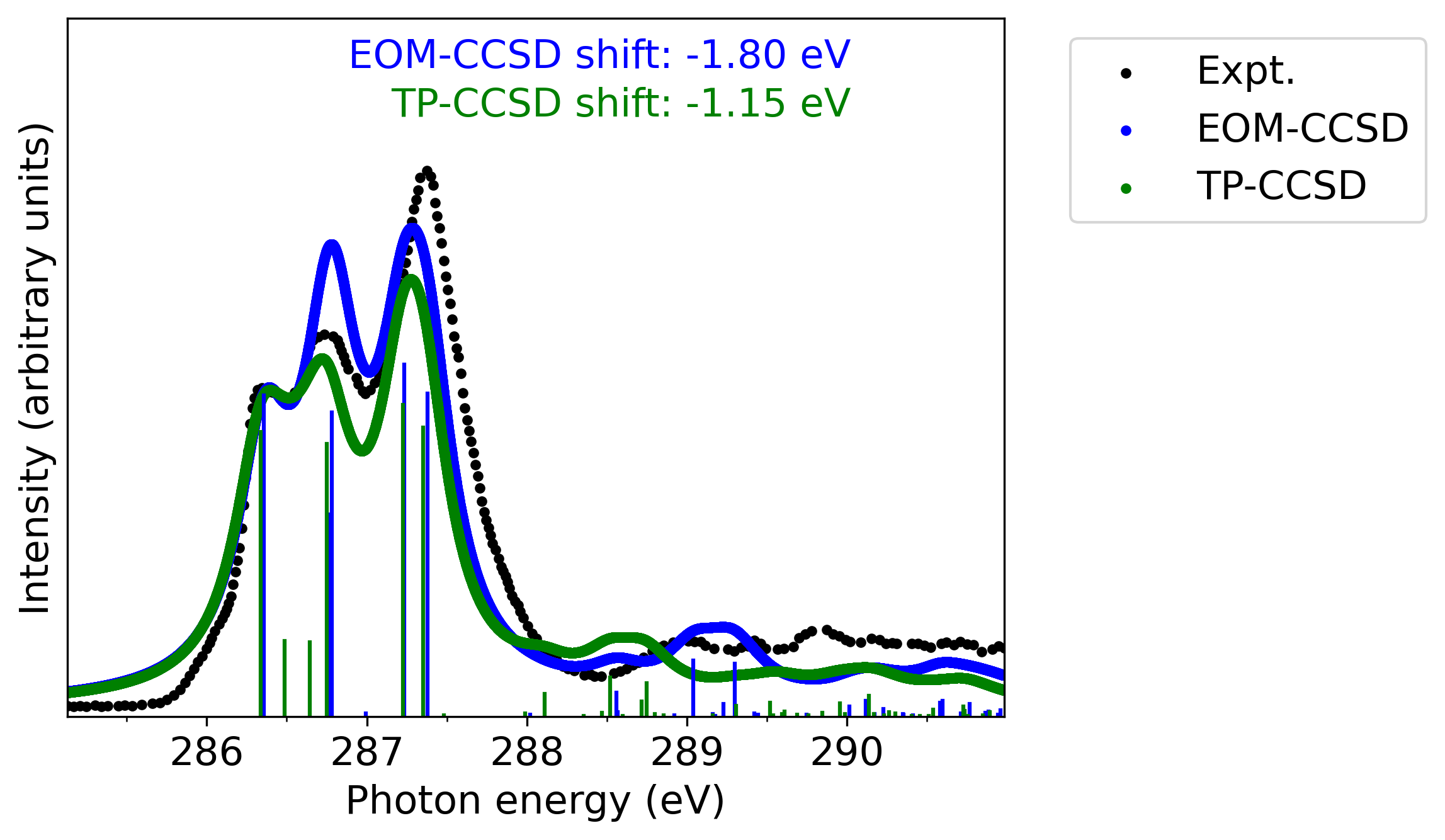}

\caption{\label{fig:adenine-kedge}Experimental and computed CVS-EOM-CCSD and
TP-CCSD K-edge absorption spectra for adenine. The experimental results
have been digitized from Ref.~\citenum{plekanTheoreticalExperimentalStudy2008}.
The computed spectra have been shifted and scaled to match the first
experimental absorption peak. Top: nitrogen K-edge, bottom: carbon
K-edge.}
\end{figure}

The N 1s XAS spectrum of adenine (\figref{adenine-kedge}, top) is
characterized by three prominent bands. The first peak at 399.5 eV
is composed of three valence transitions ($1s\rightarrow\pi^{*}$).
TP-CCSD and CCSD both reproduce this peak well, with TP-CCSD requiring
half of the shift required by EOM-CCSD. The second, broader peak at
402.1 eV arises from a number of transitions. TP-CCSD improves on
the separation between this peak and the main pre-edge peak, while
also clearly reproducing the smaller features between 401 and 401.5
eV and providing a clear and unambiguous assignment of these subtle
transitions. The third peak at 403.3 eV is of primarily Rydberg transitions,
and neither method reproduces the position of this peak well, although
TP-CCSD improves on EOM-CCSD by approximately 0.5 eV. In this region,
the insufficiency of the basis set is the main limiting factor, and
adding additional diffuse functions is necessary to more accurately
place the Rydberg transitions. The corresponding C 1s XAS spectrum
(\figref{adenine-kedge}, bottom) has a relatively simpler structure
dominated by three valence peaks. Both TP-CCSD and EOM-CCSD reproduce
the relative positions of these peaks well, although the TP-CCSD relative
intensities are distinctly improved compared to EOM-CCSD. The largest
energy change from EOM-CCSD to TP-CCSD is in the weaker transitions
appearing at $\sim286.5$ and $\sim286.6$ eV. In the EOM-CCSD spectrum,
these transitions are nearly degenerate with the stronger transitions.
Again, neither method reproduces the very weak Rydberg region of the
spectrum well. For the carbon K-edge, EOM-CCSD requires a shift of
-1.8 eV which is reduced to -1.15 eV in TP-CCSD. This reduction in
absolute energy error is more modest compared to the EOM-CCSDT benchmark.
This observation, paired with the poor performance in the Rydberg
region points to the basis set as the main limiting factor. However,
the close correlation of TP-CCSD to the full EOM-CCSDT indicates strongly
that calculations with larger basis sets could provide further improvements
over the current results. We are currently working on improving the
efficiency and scalability of our implementation in order to enable
such precise applications.

\begin{figure}
\includegraphics[scale=0.65]{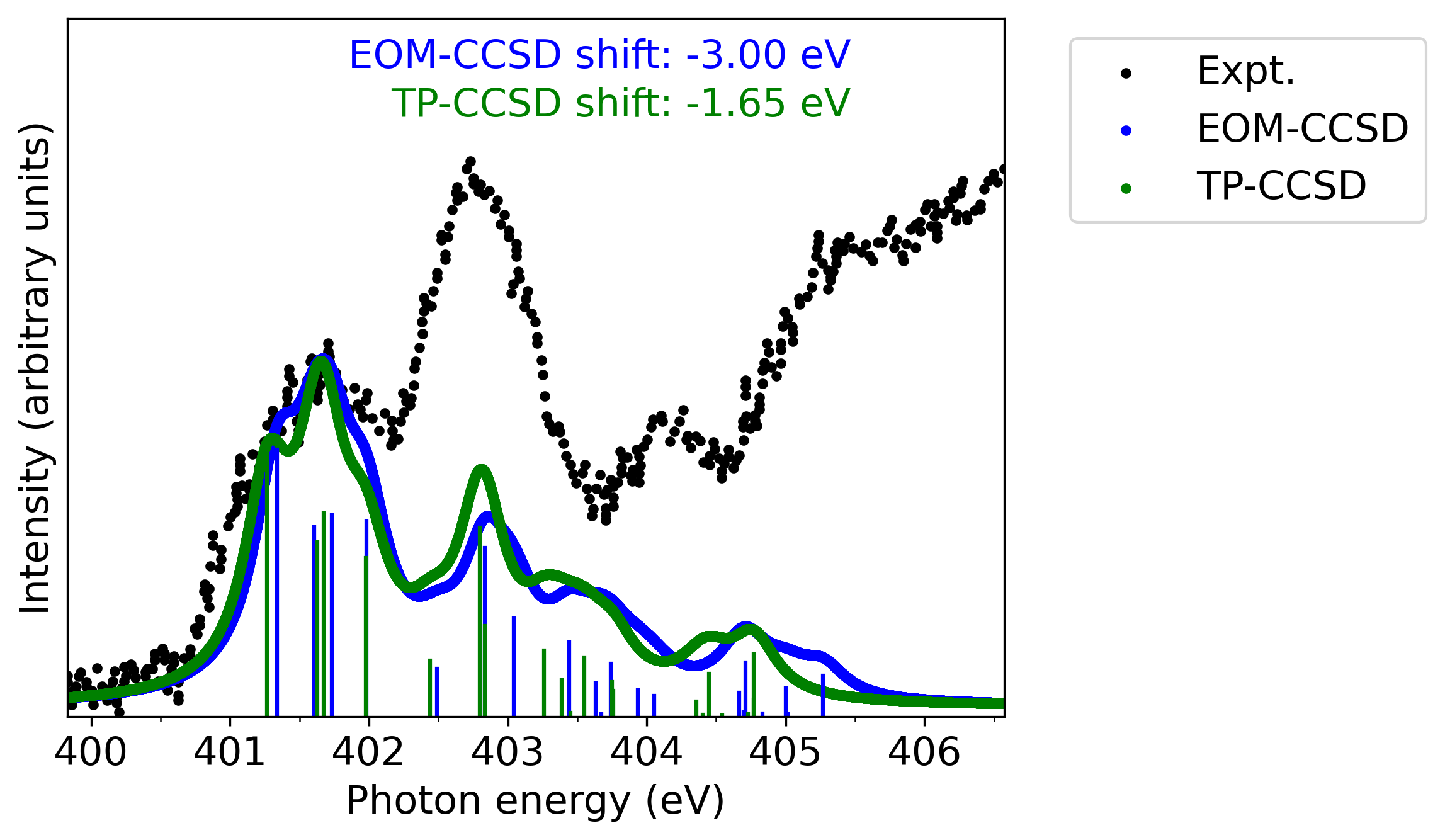}

\includegraphics[scale=0.65]{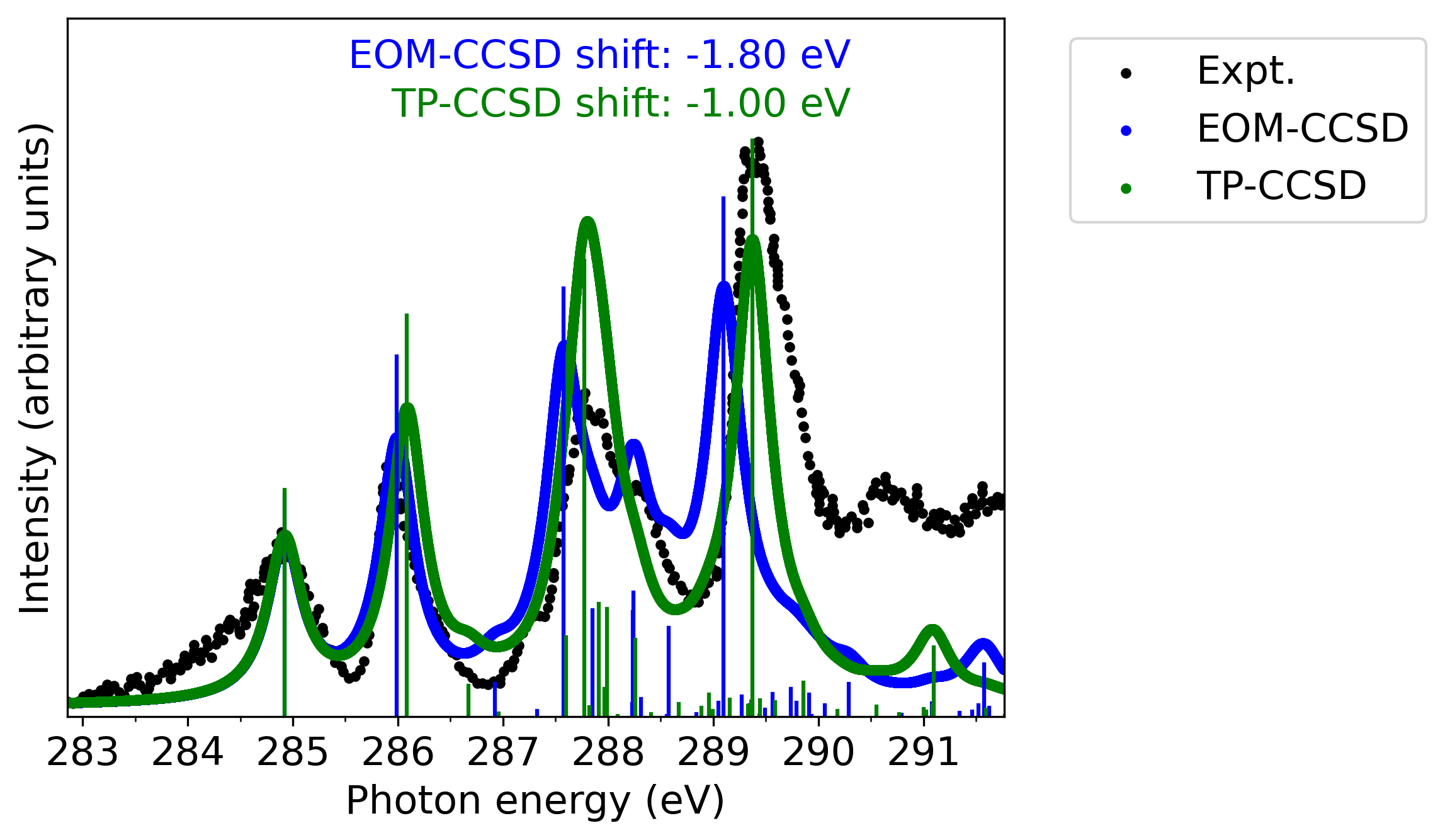}

\includegraphics[scale=0.65]{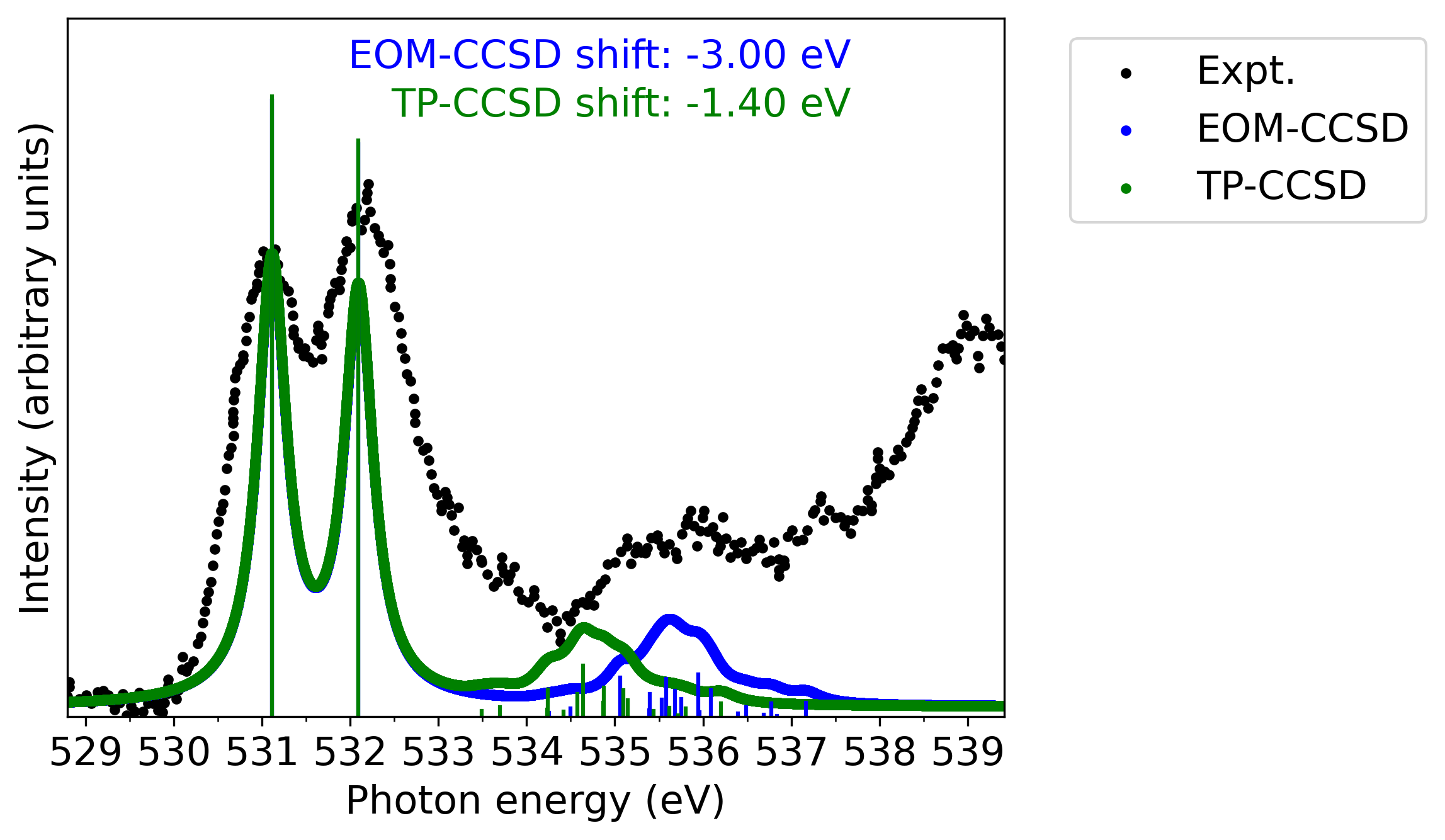}

\caption{\label{fig:thymine-kedge} Experimental and computed CVS-EOM-CCSD
and TP-CCSD K-edge absorption spectra for thymine. The experimental
results have been digitized from Ref.~\citenum{plekanTheoreticalExperimentalStudy2008}.
The computed spectra have been shifted and scaled to match the first
experimental absorption peak. Top: nitrogen K-edge, middle: carbon
K-edge, bottom: oxygen K-edge.}

\end{figure}

The nitrogen, carbon, and oxygen K-edge spectra of thymine (\figref{thymine-kedge})
show similar results, with the valence region reproduced well, the
Rydberg transitions still somewhat poorly represented due to basis
set insufficiency, and overall smaller shifts required for TP-CCSD
by about a factor of two. However, the carbon K-edge (\figref{thymine-kedge},
middle) shows some additional advantage of TP-CCSD over EOM-CCSD.
As with adenine, the spectrum shows a clear separation between intense
valence transitions and weak Rydberg transitions. However, TP-CCSD
almost exactly reproduces the positions of the four valence peaks,
spanning energies from 285 eV to almost 290 eV, while EOM-CCSD still
exhibits residual errors as high as 0.3 eV. This demonstrates the
TP-CCSD, in addition to reducing absolute energy errors and improving
relative intensities, can still provide improvements in the valence
energy structure of the spectrum compared to EOM-CCSD. 

\section{Conclusions}

Core excitation and ionization energies were calculated for a group
of small molecules using transition-potential coupled cluster {[}(X)TP-CCSD($\lambda$){]}
methods with a variety of fractional core-hole occupations (with $n_{1s;\alpha}=n_{1s;\beta}=1-\lambda/2$).
Our previous work\cite{simonsTransitionpotentialCoupledCluster2021}
showed that TP-CCSD(1/2) was as accurate for core-excited states as
EOM-CCSD is for valence states, with deviations from CVS-EOM-CCSDT
within a few tenths of an eV. In this work, we optimized $\lambda$
in an element-specific manner to identify the ``best'' core-hole
fraction for K-edges of carbon, nitrogen, oxygen, and fluorine. We
used the mean absolute error in comparison to full CVS-EOM-CCSDT and
identified clear minima in the MAE curves for both absolute and relative
energy errors, while errors in oscillator strengths did not clearly
favor a particular value of $\lambda$. We found that an element-specific
choice of the $\lambda$ parameter leads to the best accuracy (summarized
in \tabref{recommended}), although a generic value of $\lambda=0.425$
for TP-CCSD or $\lambda=0.525$ for XTP-CCSD is almost as accurate.

We then used TP-CCSD with our recommended element-specific $\lambda$
values, in combination with the more economical 6-311++G{*}{*} basis
set for calculating the C, N, and O K-edge absorption spectra of adenine
and thymine, for which gas-phase experimental data is available. TP-CCSD
was found to systematically reduce the overall energy shifts required
to match the experimental spectra, in comparison to CVS-EOM-CCSD,
while also improving the relative positions and/or intensities of
several peaks. In the Rydberg region, the insufficiency of the basis
set is the main limiting factor and adding additional diffuse functions
is necessary to increase accuracy. However, the close correlation
of TP-CCSD to the full CVS-EOM-CCSDT shows that calculations with
larger basis sets could provide further improvements over the current
results. We are currently working on improving the efficiency and
scalability of our implementation in order to enable such precise
applications.

\section*{Acknowledgments}

This work was supported in part by the US National Science Foundation
under grant OAC-2003931. MS is supported by an SMU Center for Research
Computing Graduate Fellowship. All calculations were performed on
the ManeFrame II computing system at SMU.

\section*{Disclosure Statement}

No potential conflict of interest was reported by the authors.

\bibliographystyle{unsrt}
\bibliography{paper}

\end{document}